\documentclass{article}

\usepackage{PRIMEarxiv}

\usepackage{cite}
\usepackage{amsmath,amssymb,amsfonts}

\usepackage{graphicx}
\usepackage{textcomp}
\usepackage{xcolor}
\def\BibTeX{{\rm B\kern-.05em{\sc i\kern-.025em b}\kern-.08em
    T\kern-.1667em\lower.7ex\hbox{E}\kern-.125emX}}

\usepackage{amsmath}
\usepackage{amsfonts}
\usepackage{mathrsfs}
\usepackage{amssymb}
\usepackage{dsfont}
\usepackage{times}
\usepackage{multirow}
\usepackage[none]{hyphenat}
\usepackage{float}
\usepackage{subfig}
\usepackage{iftex}
\usepackage{algorithm}
\usepackage{algpseudocode}
\usepackage{booktabs}

\newtheorem{lemma}{\bf Lemma}

\newtheorem{assumption*}{Assumption}
\newtheorem{stdassumption*}{Standing Assumption}

\newtheorem{definition}{Definition}
\newtheorem{definition*}{{Definition}}
\usepackage[acronym]{glossaries}
\newacronym{tdth}{TD-TH}{time domain thresholding}
\newacronym{tfdth}{TFD-TH}{time-frequency domain thresholding}
\newacronym{fmcw}{FMCW}{frequency modulated continuous wave}
\newacronym{pd}{PD}{probability of detection}
\newacronym{sinr}{SINR}{signal-to-interference-plus-noise ratio}
\newacronym{pri}{PRI}{pulse repetition interval}
\newacronym{adc}{ADC}{analog-to-digital converter}
\newacronym{fft}{FFT}{fast Fourier transform}
\newacronym{lrr}{LRR}{long-range radar}
\newacronym{srr}{SRR}{short-range radar}
\newacronym{cfar}{CFAR}{constant false alarm rate}
\newacronym{rfe}{RFE}{radar front end}
\newacronym{lpf}{LPF}{low-pass filter}
\newacronym{rd}{RD}{range-Doppler}
\newacronym{cdf}{CDF}{cumulative distribution function}
\newacronym{adas}{ADAS}{advanced driver assistance system}
\newacronym{ad}{AD}{automonous driving}
\newacronym{stft}{STFT}{short-time Fourier transform}
\newacronym{vru}{VRU}{vulnerable road user}
\newacronym{acc}{ACC}{adaptive cruise control}
\newacronym{tx}{TX}{transmitter}
\newacronym{rx}{RX}{receiver}
\newacronym{lfm}{LFM}{linear frequency modulated}

\title{A Game-Theoretic Approach for High-Resolution Automotive FMCW
Radar Interference Avoidance
}

\author{
  Yunian Pan,  \\
  New York University, \\ Brooklyn, \\ NY, USA \\
  \texttt{yp1170@nyu.edu} \\
   \And
 Jun Li\\
  NXP Semiconductors, \\ San Jose, \\ CA, USA \\
  \texttt{jun.li\_5@nxp.com} \\
  \AND
  Lifan Xu \\
  The University of Alabama,\\ Tuscaloosa,\\ AL, USA \\
  \texttt{lxu36@crimson.ua.edu} \\
  \And
  Shunqiao Sun \\
  The University of Alabama,\\ Tuscaloosa,\\ AL, USA \\
  \texttt{shunqiao.sun@ua.edu} \\
  \And
  Quanyan Zhu \\
 New York University, \\ Brooklyn, \\ NY, USA \\
  \texttt{qz494@nyu.edu} \\
}

\begin{document}
\maketitle

\begin{abstract}
Nonlinear frequency hopping has emerged as a promising approach for mitigating interference and enhancing range resolution in automotive FMCW radar systems. Achieving an optimal balance between high range-resolution and effective interference mitigation remains challenging, especially without centralized frequency scheduling. This paper presents a game-theoretic framework for interference avoidance, in which each radar operates as an independent player, optimizing its performance through decentralized decision-making. We examine two equilibrium concepts—Nash Equilibrium (NE) and Coarse Correlated Equilibrium (CCE)—as strategies for frequency band allocation, with CCE demonstrating particular effectiveness through regret minimization algorithms. We propose two interference avoidance algorithms: Nash Hopping, a model-based approach, and No-Regret Hopping, a model-free adaptive method. Simulation results indicate that both methods effectively reduce interference and enhance the signal-to-interference-plus-noise ratio (SINR). Notably, No-regret Hopping further optimizes frequency spectrum utilization, achieving improved range resolution compared to Nash Hopping.

\end{abstract}


\section{Introduction}

High-resolution \gls{fmcw} automotive radars have become essential to \gls{adas} and \gls{ad} due to their ability to accurately and reliably measure the range, velocity, and angle of targets in all weather conditions \cite{Sun_SPM_2020,Sun_JSTSP_4D_2021,Lifan_Co_Chirps_TAES_2023,Ruxin_Radar_Imaging_TAES_2023,Markel_book_2022,BLRC}. 
However, the rapid increase in the deployment of automotive radars has led to a growing concern over radar-to-radar interference, where unwanted signals from nearby radars degrade the quality of target detections.
To address this, various interference mitigation techniques \cite{li2024performance,jun2022radar,jeroen19,Mun18,Mun20,jihwan2024intf,xinyi2024} have been proposed, aimed at enhancing the safety and comfort functions of the automotive radar systems. Depending on the specific automotive radar functions, the necessary improvement in the \gls{sinr} after interference mitigation varies from a few dB up to infinite. For instance, in radar-based \gls{vru} protection applications, any sudden system failure caused by interference is unacceptable, necessitating complete interference avoidance in such critical situations.

Proactive interference mitigation on the \gls{tx} side \cite{Lifan_Co_Chirps_TAES_2023,Sun_JSTSP_4D_2021,Yimin_Partioning_SAM_2024} is among the most effective techniques for avoiding radar interference. These methods can completely suppress interference by transmitting signals in time slots or subbands where no interference signals are present, thereby ensuring interference-free operation for safety-critical functions, e.g., \gls{acc}, collision mitigation, and \gls{vru} protection. For instance, in \cite{stettiner2023fmcw}, a technique known as frequency hopping is introduced to mitigate radar interference by rapidly switching the carrier's starting frequency across a wide range of available subbands. Thus, the likelihood of collisions is reduced. However, the interference problem still prevails as the FMCW radar channel parameters, such as frequency and time offsets, are often randomly assigned without coordination. This lack of structured frequency assignment makes frequency hopping alone insufficient for interference avoidance. Additionally, it limits adaptability in dynamic situations.  
While improvements can be achieved through techniques like minimum variance beamforming\cite{Jin_Interference_TVT_2024}, the primary issue stems from improper frequency scheduling. Other proposed solutions involve using communication networks, such as high-bandwidth LTE links, to form radar networks and adopting Medium Access Control (MAC) protocols. 
These approaches, whether centralized \cite{Google_Automotive_Interference_Mitigation_2016} or distributed \cite{Jin_FMCW_Interference_JSTSP_2021}, rely heavily on either a cloud infrastructure or side communication channels, and can be vulnerable to adversarial disruptions. 



To develop a more adaptive solution for frequency scheduling, we resort to game theory, modeling radars as players with performance metrics and frequency hopping schemes as their utilities and strategies.
In noncooperative game theory, the most common notion is \textit{Nash Equilibrium} (NE) \cite{bacsar1998dynamic}, where players have no incentive to deviate from their independent strategies.
Another competing notion is \textit{Coarse Correlated Equilibrium} (CCE) \cite{roughgarden2010algorithmic}, which are the ``optimal'' joint distributions over the players' hopping strategies, in the sense that the players have no incentive to deviate from the frequency decisions sampled from a CCE.

The concept of NE inherently supports a distributed, incentive-compatible, and stable frequency-hopping solution that may potentially lead to a \textit{system optimum} (SO).
However, it presents several limitations in the radar communication scenarios. First, NE is prone to equilibrium selection issues, as players choosing different NE strategies can result in non-equilibrium outcomes; Second, the average radar performance achieved by an NE is not guaranteed to match that of an SO; Third, finding an NE requires either complete knowledge of the environment and other radars-typically inaccessible and computationally intensive (a PPAD-complete problem) \cite{thecomplexityofne}-or a learning process with repeated interactions, which does not guarantee convergence. Conversely, convergence to a CCE is more achievable through interactive learning due to the diminishing regrets of individual radars—the analysis of which is well-documented in the literature (e.g., \cite{pan2023resilience, pan2024variationalinterpretationmirrorplay}). 
For this reason, we adopt CCE as the primary solution concept and propose \textit{No-regret Hopping}, an effective distributed frequency scheduling method for avoiding radar interference. Then we compare the \textit{No-regret Hopping} method with a centralized method \textit{Nash Hopping} \cite{9682998}, which targets an NE through model-based learning and also uses the uniformly random strategy as another experimental benchmark. Our simulations indicate that radars with \textit{No-regret Hopping} achieve superior \gls{sinr} and lower interference rates in interference conditions while improving range resolution by utilizing the available interference-free bandwidth more effectively.


\section{Problem Formulation}
\label{sec:radarsystemasgame}

Consider a scenario where a set of FMCW automotive radars, denoted by $\mathcal{I}:=\{1, \ldots, I\}$, experience mutual interference. Each radar transmits \gls{lfm} pulses within a total bandwidth $B$ and starting frequency $f_c$. The nonlinear frequency hopping strategy \cite{stettiner2023fmcw} divides the bandwidth $B$ into $A$ subbands, $\mathcal{A}:=\{ f_1,\ldots, f_A\}$, with each subband defined by a corresponding starting frequency $f_a = f_c + (a-1)B_a$ for $a \in \{1, \ldots, A\}$, and each chirp only sweeps a portion $B_a$ of the total bandwidth $B$.

\subsection{The Anti-Coordination Game}
Consider a time frame with total duration $T$, 
where each radar $i \in \mathcal{I}$ transmits $K^i$ consecutive chirps. The pulse repetition interval (PRI) is $T^i_{\text{PRI}} = T/K^i$, encompassing both active and idle times, where $(T^i_{\text{PRI}} = T^i_a + T^i_d)$ and $T^i_d > 0$. This results in a chirp sweeping slope of $\alpha^i = B_a/T^i_a$ for radar $i$. For simplicity, we assume $K^i = K$ for all $i$. To minimize interference, the radars strategically select channel sequences $(f^i_1, \ldots, f^i_K)$ across the chirps.

Then, the radar-to-radar interference scenario can be modeled as a $K$-stage repeated game $\mathcal{G} = \{ \mathcal{I}, \mathcal{A}, \{ U_i \}_{i \in \mathcal{I}}\}$, where the radars, acting as players in $\mathcal{I}$, share a common action set $\mathcal{A}$. The anti-coordination nature lies in the utility functions $U_i: \mathcal{A}^{I} \to \mathbb{R}$, where $f^i, f^{-i} \in \mathcal{A}$ denote the subbands chosen by player $i \in \mathcal{I}$ and by all other players, respectively. In this paper, the utility function for player $i$ is defined as:
\begin{equation}\label{eq:utility}
    U_i (f^i, f^{-i}) =   10\log_{10}(\mathrm{SINR} (f^i, f^{-i})) ,
\end{equation}
where $\mathrm{SINR}(f^i, f^{-i})$ represents the \gls{sinr} of subband $[f^i, f^i+B_a]$ at radar $i$'s receiver, based on the selected subband actions $f^i$ and $f^{-i}$. In the absence of interference, the SINR reduces to the signal-to-noise ratio (SNR). If any starting frequency in $f^{-i}$ matches $f^i$, the $U_i$ decreases, highlighting the anti-coordination objective, as radars aim to avoid overlapping frequencies to maintain higher utility.

\subsection{Signal Model}

Let radar $i$'s starting frequency sequence be $(f^i_{1},\cdots,f^i_{K} )$, and consider $\mathcal{N}:=\{1,\ldots,N\}$ targets within its field of view, with range vector $(r^i_1, \ldots, r^i_N)$ and velocity vector $(\dot{r}^i_1, \ldots, \dot{r}^i_N)$. The velocities are negative for approaching targets.
The transmitted signal at the $k$-th chirp can be expressed as:
\begin{equation}\label{eq:transmittedsignal}
    s^i[t, k]=e^{j 2 \pi\left[f^i_k t+\frac{1}{2} \alpha^i t^2\right]}, \quad 0\leq t<T_a^i.
\end{equation}
The received signal reflected from target $n \in \mathcal{N}$, after passing through the mixer, can be written as:
\begin{equation} \label{eq:echosignal}
    \begin{aligned}
 y^i_n[t, k] &= a^i_n  s^i\left[t-\Delta^i_n t_k, k\right] s^{i*}[t, k] \\
& \approx a^i_n e^{ -j 2 \pi\left[f^i_k \Delta^i_n{t_k}+\alpha^i \Delta^i_n{t_k} t\right]},
\end{aligned}
\end{equation}
where $a^i_n$ is the complex target coefficient. $\Delta^i_n{t_k}$ denotes the round trip delay of $k$-th chirp reflected from the target and can be approximated by $\Delta^i_n{t_k}=\frac{2}{c}\left(r^i_n +k \dot{r}^i_n T^i_{\text{PRI}}\right)$, where $c$ is the speed of light.

The true range value can be expressed as
\begin{align}
    r^i_n  +k \dot{r}^i_n T^i_{\text{PRI}}=\bar{r}^i_n+\left(\epsilon_0+k \dot{r}^i_n T^i_{\text{PRI}}\right),
\end{align}
where $ \bar{r}^i_n$ represents the target's coarse range center, with a bin width of $\frac{c}{2B_a}$. The fine quantization $\epsilon_0 \in\left[-\frac{c}{4 B_a}, \frac{c}{4 B_a}\right]$ has a bin width of $\frac{c}{2 B}$. Assuming negligible range migration for the coarse range \cite{stettiner2023fmcw}, the received target signal in \eqref{eq:echosignal} can be rewritten as
\begin{equation}
\begin{aligned}
    y^i_n [t, k] = \tilde{a}^i_n  e^{-j2\pi f_rt} e^{j2 \pi f_dk} e^{-j2\pi [\frac{2\bar{r}^i_n}{c} + \frac{2(\epsilon_0+k \dot{r}^i_n T^i_{\text{PRI}} )}{c}]\Delta b^i_k },
    \label{eq:finerange}
\end{aligned}
\end{equation}
where constant terms are consolidated in the complex target coefficient $\tilde{a}^i_n$. Here, $f_r=\frac{2r^i_n\alpha^i}{c}$ is the coarse range frequency, and $f_d=-\frac{2\dot{r}^i_n T^i_{\text{PRI}}f_c}{c}$ is the Doppler frequency. The last term represents the phase compensation across slow time due to frequency hopping, where $\Delta b^i_k=f^i_k-f_c$ is the known frequency shift at both the transmitter and receiver. Without frequency hopping, $\Delta b^i_k=0$, and a conventional received signal remains; with frequency hopping, the last phase compensation term in \eqref{eq:finerange} allows for higher range resolution. After the received signal is processed through a low-pass filter (LPF) and sampled by an analog-to-digital converter (ADC), the range FFT over index $t$ extracts the coarse range information. Then, applying a modified Doppler FFT over index $k$ and nonlinear frequency hopping sequence $\Delta b^i_k$ enables extraction of both velocity and fine range information \cite{stettiner2023fmcw}.

At radar $i$'s receiver, assume that the received target signal is mixed with signals directly transmitted by other radars $ \texttt{o} \in  \mathcal{I}/\{i\}$ with carrier frequencies $f^{\texttt{o}}_k$. If $f^{\texttt{o}}_k=f^{ i }_k$, the radar-to-radar interference occurs. The demodulated and dechirped interference signal at radar $i$'s receiver
can be expressed as  
\begin{equation}
\begin{aligned}
     \quad  y^{\texttt{o}}_n [t, k] &= a^{\texttt{o}}_n s^{\texttt{o}}[t-\Delta^{\texttt{o}}_n t_{k}, k] s^{i *}[t, k]  \\ 
    & \approx \tilde{a}^{\texttt{o}}_n   e^{j\pi(\alpha^i -\alpha^{\texttt{o}})t^2 }, 
\end{aligned}
\end{equation}
where $a^{\texttt{o}}_n$ is the complex interference coefficient and $\alpha^{\texttt{o}}$ is the chirp slope of radar $\texttt{o}$. Thus, the dechirped interference signal becomes a new chirp signal with a slope of $\alpha^i -\alpha^{\texttt{o}}$. In practice, the start time and frequency of interference depend on factors such as the chirp parameters of both the victim and interfering radars, the position of the interfering radar, and other conditions \cite{li2024performance}. 

By combining the target signal and the interference signal, the received signal at radar $i$ can be expressed as:
\begin{equation}
     \mathbf{\hat{y}}^i_k = \mathbf{y}^i_{k}+\mathbf{y}^{\texttt{o}}_{k} + \mathbf{e}^i_k , 
\end{equation}
where $\mathbf{y}^i_{k}$ and $\mathbf{y}^{\texttt{o}}_{k}$ are the low-pass filtered ADC samples of $\sum_{n \in \mathcal{N}}y^i_n[t,k]$ and $\sum_{\texttt{o} \in \mathcal{I}/\{i\}, n \in \mathcal{N}}y^{\texttt{o}}_n [t,k]$ within chirp $k$, respectively, and $\mathbf{e}^i_k$ represents the white Gaussian noise.

Let $\mathrm{P}(\cdot)$ denote the average power function of a signal source. The theoretical $\mathrm{SINR} (f^i_k, f^{-i}_k)$ at radar $i$'s receiver can be expressed as
\begin{equation}\label{eq:theoreticalSNR}
    \mathrm{SINR} (f^i_k, f^{-i}_k)= \frac{\mathrm{P}(\mathbf{y}^i_{k})}{  \mathrm{P}(\mathbf{y}^{\texttt{o}}_{k}) + \mathrm{P}(\mathbf{e}^i_k)}.
\end{equation} 
In practice, it requires detection techniques to determine the existence of interference, e.g., the thresholding-based strategy \cite{li2024performance}, which filters the received signal $\mathbf{\hat{y}}^i_k$ into an interference component $\mathbf{\tilde{y}}^{\texttt{o}}_{k}$ and a clean signal $\mathbf{\tilde{y}}^i_k$ after interference mitigation. However, due to the stochastic nature of the signals and the lack of visibility into other radars' subband choices, we use a sliding window approach to estimate SNR and SINR periodically for each subband. The entire hopping process is divided into episodes, with $\boldsymbol{\tau} := \{k_{\tau-1} + 1, \ldots, k_{\tau}\}$ as the $\tau^{th}$ episode, where $\tau=1,\ldots, \mathcal{T}$, $  K^i \equiv 0   \pmod{\mathcal{T}} $, and $k_{\tau} = \tau K^i / \mathcal{T}$. 
For each episode $\tau$ at radar $i$, we estimate the SINR for each subband $f$ as:
\begin{equation}
\begin{aligned}
    \label{eq:estima}
     \overline{\mathrm{SINR}}_{\tau} (f)  =  \frac{1}{\sum_{k\in \boldsymbol{\tau} }\mathds{1}\{ f^i_k = f\}}\sum_{k\in \boldsymbol{\tau} }\frac{\mathrm{P}(\mathbf{\tilde{y}}^i_k ) \mathds{1}\{ f^i_k = f\}}{\mathrm{P}(\mathbf{\tilde{y}}^{\texttt{o}}_{k}) + \mathrm{P}(\mathbf{e}^i_k)} ,
\end{aligned}
\end{equation}
whenever interference is detected. When interference-free chirps are present, the SNR can be estimated as: $
     \overline{\mathrm{SNR}}_{\tau} (f)  =   \frac{1}{\sum_{k\in \boldsymbol{\tau} }\mathds{1}\{ f^i_k = f, \mathbf{\tilde{y}}^i_k = \mathbf{y}^i_k\}}\sum_{k \in \boldsymbol{\tau}}\frac{\mathrm{P}(\mathbf{y}^i_{k} )}{ \mathrm{P}(\mathbf{e}^i_k)} \mathds{1}\{ f^i_k = f, \mathbf{\tilde{y}}^i_k = \mathbf{y}^i_k\}.$ 


\subsection{Frequency Hopping with Mixed Strategy}

If each radar occupies a unique subband, no interference occurs across the chirps. However, with multiple subbands available in frequency hopping, only using a single unique subband degrades range resolution, as range resolution is inversely proportional to the total bandwidth utilized. To address this trade-off, we equip each radar $i$ with an adaptable mixed strategy $p^i_{\tau}$ for each episode $\tau = 1,\ldots,\mathcal{T}$. This strategy is a probability distribution over $\mathcal{A}$, i.e., $p^i_{\tau} \in \mathcal{P}(\mathcal{A}) := \{ p: \mathcal{A} \to [0,1]\  \big| \ \sum_{f \in \mathcal{A}} p(f) = 1 \} $ for all $f \in \mathcal{A}$. Before the start of each chirp $k \in \boldsymbol{\tau}$, each radar samples a starting frequency decision, which we will write as $f^i_k \sim p^i_\tau(\cdot)$ throughout the rest of the paper without explicitly identifying $\tau$. In essence, the goal is for radars to employ mixed strategies to generate a sequence of transmitted signals that effectively detect moving targets while balancing the trade-off between range resolution and interference mitigation.


\section{Solution Concepts and Algorithms}\label{sec:solconceptalgo}

In this section, we present two game-theoretic solution concepts for the anti-coordination game, along with corresponding algorithmic approaches to achieve these solutions during the frequency hopping process. 

\subsection{Nash Equilibrium and Coarse Correlated Equilibrium}
 
Let $p = (p^i, p^{-i})$ be a mixed strategy profile, it constitutes an NE during every chirp $k$ if no radar has the incentive to deviate from their own mixed strategy, in the sense of that the expected utility, 
denoted by $\bar{U}_i: \mathcal{P}(\mathcal{A}^I) \to \mathbb{R}$, 
\begin{equation}
    \bar{U}_i (p^i \otimes p^{-i}):=  \mathbb{E}_{ f^i \sim p^i, f^{-i} \sim p^{-i}} [ U_i (f^i, f^{-i})] ,
\end{equation}
where $p^i \otimes p^{-i}$ is the outer product of the players' mixed strategies, cannot be improved. 

\begin{definition}[Nash Equilibrium (NE)]
 For the $K$-stage repeated game $\mathcal{G}$, a mixed strategy profile $(p^{i,\star}, p^{-i,\star})$ constitutes a Nash Equilibrium (NE) if for all radars $i \in \mathcal{I}$, and all other strategies $p^i \in \mathcal{P} (\mathcal{A})$, the following inequalities are satisfied: 
 \begin{equation}
      \bar{U}_i (p^{i,\star} \otimes p^{-i,\star}) \geq \bar{U}_i (p^{i} \otimes p^{-i,\star}) .
 \end{equation}
  We denote by $\mathscr{N}(U)$ the set of NE mixed strategy profiles with $U$ being the utility functions.
 \end{definition}


Due to the lack of information about $U$, we adopt an \textit{explore-then-commit} type of strategy in Algorithm \ref{algo:nashtransmit}. 
During \texttt{Nash Hopping}, we partition the chirping horizon $K$ into an exploration phase $k = 1, \ldots, k_{e}$, where the goal is to estimate the channel capacity functions $U_i$ through uniform mixed strategies $p_e^i$; and a commitment phase $k = k_e +1, \ldots, K$, where the radars are committed to a mixed strategy Nash equilibrium that maximizes the social welfare.

\begin{algorithm}[htbp]
\caption{\texttt{Nash Hopping}}\label{algo:nashtransmit}
\begin{algorithmic}[1]
\State \textbf{Input:} 
Initialize $p^i_e = \mathrm{Uniform}(\mathcal{A})$ for all radars $i \in \mathcal{I}$.
\State \textbf{Output:}
Mixed strategies for interference avoidance

   \For{all $i \in \mathcal{I}$ in parallel}
   set $\tau = 1$.
   \For{exploration chirps $k = 1:k_e $}
   \State  Sample $f^i_k \sim p^i_{e}(\cdot)$ for all radars $i \in \mathcal{I}$
     \If{$ k = k_{\tau}$}
     \State 
       Assign utilities:
      $$
     \begin{aligned}
         \tilde{U}_i (f^i, f^{-i} ) \leftarrow 10
    \log_{10}(\overline{\mathrm{SINR}}^i_{\tau}(f^i) )  \exists j \in \mathcal{I} /\{i\} f^i = f^{j}  \\
     \tilde{U}_i (f^i, f^{-i} ) \leftarrow 10 
    \log_{10}(\overline{\mathrm{SNR}}^i_{\tau}(f^i) )  \forall j \in \mathcal{I} /\{i\} f^i \neq f^{j} 
     \end{aligned}$$ 
     \State $(p^i_{e}, p^{-i}_{e}) \gets \arg\max_{ p \in \mathscr{N}(\tilde{U})} \sum_{i \in \mathcal{I}} \tilde{U}_i $ 
     \State  Update $\tau = \tau +1 $ 
\EndIf
\EndFor
\EndFor

\For{committing chirps $k= k_e +1 : K$}
\State Sample $f^i_k \sim p^i_e(\cdot)$ for all $i \in \mathcal{I}$ 
\EndFor
\end{algorithmic}
\end{algorithm}

\texttt{Nash Hopping} requires information exchange despite that estimating $\bar{U}_i$ requires only local interference detection and SINR computation; to compute an (approximated) NE, one needs the estimated utility functions $\tilde{U}_i$ for all radars $i$. This drawback limits its application only to scenarios where information sharing is possible. Besides, achieving an equilibrium state necessitates the coordination of all radars, as the uniqueness of the NE is not guaranteed. Nevertheless, the computational complexity of NE grows exponentially in the number of radars and subbands.

\begin{definition}
     For the $K$-stage repeated game $\mathcal{G}$, a joint probability distribution $ \pi \in \mathcal{P}(\mathcal{A}^I)$ is said to be a Coarse Correlated Equilibrium (CCE) if the following equalities are satisfied: 
    \begin{equation}\label{eq:ccebimatrix}
            \mathbb{E}_{f \sim \pi } [ U_i ( f^i, f^{-i} ) ] \geq \mathbb{E}_{ f \sim \pi }[  U_i(f^{i,\prime}, f^{-i})] ,
    \end{equation}
    for all $f^{i,\prime} \in \mathcal{A}$. The interpretation is that whenever radar $i$ deviates from a joint decision $f$ sampled from a CCE $\pi$ by choosing some subband $f^{i,\prime}$, its expected utility cannot be improved.
\end{definition}

Compared to NE, Coarse Correlated Equilibrium (CCE) allows for decentralized solutions. By correlating individual hopping strategies, CCE facilitates coordination, leading to more efficient frequency selection. However, achieving CCE requires a common signaling mechanism, which can be addressed using regret-minimization-based learning.

\subsection{Regret Minimization and Implicit Regularization }\label{sec:regretminregularize}

Suppose an interference mitigation scheme generates a sequence $\{f^i_k, f^{-i}_k\}_{k=1}^K$. Then, 
the (external) regret of hopping process for radar $i$ is, as defined in \eqref{eq:extregret}, 
\begin{equation}\label{eq:extregret}
     \mathcal{R}_i =  \max_{f^i \in \mathcal{A}}  \sum_{k=1}^{K}  [ U_i (f^i, f^{-i}_k ) - U_i ( f^i_k, f^{-i}_k)] .
\end{equation}
This regret notion captures the difference between decisions made and the hindsight optimal.  A regret-minimization algorithm allows radars to adapt their frequency hopping strategies over time while minimizing long-term interference.

\begin{algorithm}
\caption{\texttt{No-Regret Hopping}}\label{algo:noregrettransmit}
\begin{algorithmic}[1]
\State \textbf{Input:}
For all radars $i \in \mathcal{I}$, initialize $p^i_1 = \mathrm{Uniform}(\mathcal{A})$, 
exploration parameter $\gamma_{\tau}$,
learning rate $\eta_{\tau}$, and set\\
loss vector $ \hat{L}^i_{0, f} = 0$ for all $f \in \mathcal{A}$, $i \in \mathcal{I}$.
\State \textbf{Output:}
Mixed strategies for interference avoidance
\For{all $i \in \mathcal{I}$ in parallel}
set $\tau = 1$,
\For{$k = 1:K^i$}

    \State Sample subband  $ f^i_k \sim  p_{\tau}^i (\cdot)$ 
    
    \If{$ k = k_{\tau}$}
    \State Update $\hat{L}^i_{\tau} $, for all $f \in \mathcal{A}$: 
    \begin{equation}
          \hat{L}^i_{\tau,f} = \hat{L}^i_{\tau-1, f} - \frac{10\log_{10}(\overline{\mathrm{SINR}}_{\tau}^i  (f) ) }{ p^i_{\tau}(f) } 
    \end{equation}
    \State Calculating mixed strategy: 
    \begin{equation}
    \begin{aligned}
      \tilde{p}^i_{\tau+1} (\cdot) & = \left(\frac{ \exp( -\eta_{\tau} \hat{L}_{\tau, f} )}{\sum_{f^{\prime} \in \mathcal{A}} \exp( - \eta_{\tau}  \hat{L}_{\tau, f^{\prime}} )} \right)_{f \in \mathcal{A}} \\
        p^i_{\tau+1} (\cdot) & =  (1 - \gamma_{\tau}) \tilde{p}^i_{\tau+1}(\cdot) + \gamma_\tau \mathrm{Uniform} (\mathcal{A})
    \end{aligned}
    \end{equation}
    \State Update $\tau = \tau + 1$
     \EndIf 
    \EndFor
\EndFor

\end{algorithmic}
\end{algorithm}
Algorithm \ref{algo:noregrettransmit} \texttt{No-Regret Hopping} is an epitome of such methods. 
It ensures that the regret for radar $i$ remains small, regardless of the decision-making of other radars, and concentrates around its expected value.
In the meantime, the exploration parameter $\gamma_{\tau} \geq 0$ introduces optimistic biases over the less favored frequency bands, this implicit regularization smooths the distribution $p^i_{\tau}$ so that the subbands with lower $\mathrm{SINR}$'s will still be chosen, avoiding converging too fast to the pure strategy. 



\subsection{Connection to Coarse Correlated Equilibrium}

We hereby discuss the performance guarantee from a theoretical point of view.
We start with a well-established result (Lemma \ref{lemma:ccenoregret}) from algorithmic game theory \cite{roughgarden2010algorithmic} that bridges the regret analysis and the CCE. Let the empirical joint distribution of transmission strategies be 
\begin{equation}\label{eq:emp}
     \overline{\pi} ( f^i, f^{-i}):= \frac{ \sum_{k=1}^{K} \mathds{1}\{ f^i_k = f^i, f^{-i}_k = f^{-i}\}}{K}
\end{equation}
for all $(f^i, f^{-i}) \in \mathcal{A}^I$.

\begin{lemma}\label{lemma:ccenoregret}
    During the hopping process, the empirical joint distribution of frequency hopping strategies $\bar{\pi} $ within chirp horizon $K$ is a $\varepsilon$-CCE, i.e.,
    \begin{equation}
         \mathbb{E}_{f \sim \bar{\pi}} [ U_i( f^i, f^{-i} ) ] \geq \mathbb{E}_{ f \sim \bar{\pi} }[  U_i(f^{\prime}, f^{-i})] - \varepsilon , 
    \end{equation}
    for all $f^{\prime} \in \mathcal{A}$, 
    if all radars follow the $\varepsilon$-no-regret learning dynamics, i.e., $\frac{1}{K} \mathbb{E}[\mathcal{R}_i] \leq \varepsilon$ for all $i \in \mathcal{I}$. 
\end{lemma}

Let $\mathcal{O}(\cdot)$ be the big-O notation, we present an informal result regarding the performance quantification of  Algorithm \ref{algo:noregrettransmit}. 
That is, if each radar $i$ is following \texttt{No-regret Hopping}, with  parameters $\eta_{\tau}, \gamma_\tau = \mathcal{O}(\sqrt{\frac{\log(A)}{ \tau A}})$, their external regret satisfies
      $ \mathbb{E} [\frac{1}{\mathcal{T} }   \mathcal{R}_i ] \leq \mathcal{O} \left(\sqrt{\frac{ A \log A}{\mathcal{T}}} \right).$ 
The corollary is that when $K$ is scaled by $\mathcal{T}$ through a constant, the empirical joint distribution $\bar{\pi}$ is an $\mathcal{O} \left(\sqrt{\frac{ A \log A}{\mathcal{T}}} \right)$-CCE, which implies the convergence:
if we let $\mathcal{T}  \to \infty $, the empirical distribution gets arbitrarily close to a CCE. 

Note that this result is also coupled with a high-probability version, interested readers can refer to online-learning literature such as \cite{lattimore2020bandit} for formal statements and proofs. 
 In fact, in our specific radar anti-coordination game scenario, the simulations show that in two radar cases, not only the empirical distribution, but the last-iterate joint distribution also converges to the set of CCE.

\section{Numerical Results}
In this section, we examine an interference scenario where two vehicles, each equipped with an FMCW radar (Radar $1$ and Radar $2$), are approaching each other. Both radars employ frequency hopping and apply the proposed game-theoretic frameworks, \texttt{Nash Hopping} and \texttt{No-Regret Hopping}, for subband hopping scheduling. The radar parameters are summarized in Table.\ref{para}.

\begin{table}[htbp] 
\centering
\caption{Radar Configuration}\label{para}
\begin{tabular}{lcc}
    \toprule
    \textbf{Parameter}              & \textbf{Radar 1} & \textbf{Radar 2} \\
    \midrule
    Sub-bandwidth (MHz)  $B_a$               & 150              & 150             \\
    Number of subbands $|\mathcal{A}|$ & 6 & 6\\
    Carrier starting frequency (GHz) $f_c$        &  77               & 77              \\ 
    Pulse repetition interval ($\mu s$) $T^i_{\text{PRI}}$  & 20            &   40       \\
    Target range ($m$)  $r^i$             & 20              & 20             \\
    Target velocity ($m/s$)  $\dot{r}^i$         & -15             & -15             \\
    Number of chirps per frame $K^i$  &  512  &  256 \\
    TX power (dBm) $P_{tx}$ &  13 &  13\\
     \bottomrule
\end{tabular}
\end{table}
We consider a challenging interference scenario in which both radars are initially synchronized, with their starting time offsets set to zero. Unlike the theoretical framework, we allow the radars to operate with different PRIs, resulting in a differing number of chirps per frame. However, both radars operate on the same CPI time scale; therefore, Radar $1$ updates its mixed strategy every $512$ chirps, while Radar $2$ does so every $256$ chirps. Although this synchronization is enforced for convenience, it is not strictly necessary. As long as the radars gather sufficient information within successive time frames, the strategy update can be unsynchronized.
\begin{figure*}[htbp] 
    \centering
        \includegraphics[width=.495\textwidth]{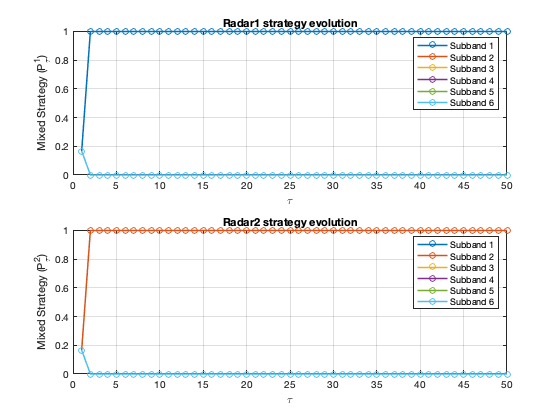}
        \includegraphics[width=.495\textwidth]{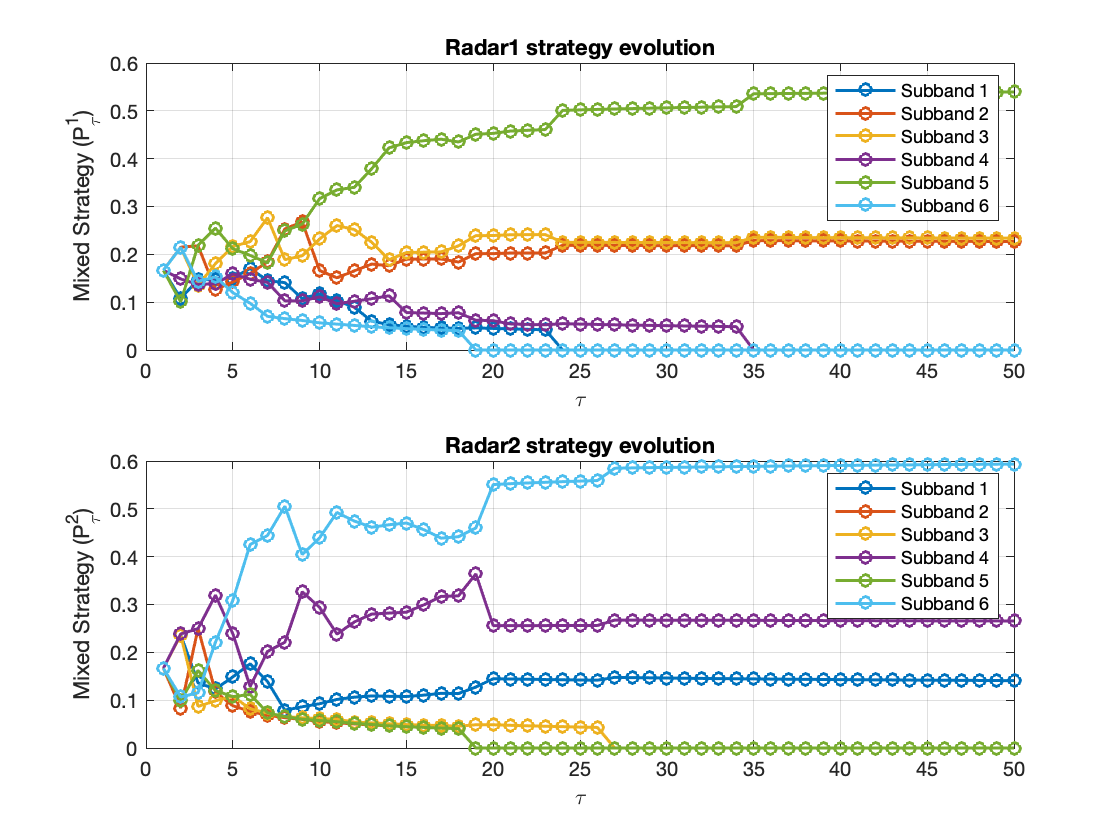}
        \label{fig:nashandnoreg}
    \caption{The evolution of mixed strategies 
    for \texttt{Nash Hopping} (left),  and \texttt{No-Regret Hopping} (right).
}
    \label{fig:main1}
\end{figure*}

We assume a uniform radar setup, meaning that the algorithms implemented in each radar are identical. Then, we evaluate the two game-theoretic methods, \texttt{No-Regret Hopping} and \texttt{Nash Hopping}, in a mutual interference scenario
and plot the evolutions of the mixed strategies $p^1_{\tau} (\cdot), p^2_\tau (\cdot)$ for these two methods over $50$ episodes.
In our experiment, we adjusted the episode length to a single time frame—a slight deviation from the approach described in Section \ref{sec:radarsystemasgame}-which enabled us to obtain more accurate utility estimates during the exploration phase. In the \texttt{No-Regret Hopping} method, we apply ``hard thresholding'' to the mixed strategy vectors, setting $p^i_\tau(f) = 0$ for subbands $f \in \mathcal{A}$ where $p^i_\tau (f) \leq \kappa$, with $\kappa$ being a small positive threshold ($\kappa = 0.04$ in our experiment). The probability vector is then renormalized. For \texttt{Nash Hopping}, we consider an exploration phase of $10$ episodes, i.e., $10\times K^i$ for radar $i = 1, 2$. 

As shown in Fig. \ref{fig:main1}, \texttt{Nash Hopping} effectively eliminates interference after the initial exploration phase. 
We coordinate two radars to adopt the strategies from one of the multiple pure Nash equilibria, where their signal transmission schemes concentrate on two distinct subbands. 
In contrast, \texttt{No-Regret Hopping} produces smoother learning curves, with each radar’s mixed strategy consistently supported by disjoint sets of three subbands. From an interference avoidance perspective, both \texttt{Nash Hopping} and \texttt{No-Regret Hopping} successfully mitigate interference by dynamically scheduling frequency hopping into interference-free subbands. However, \texttt{Nash Hopping} requires inter-radar communication as discussed in Section III, whereas \texttt{No-Regret Hopping} operates independently, without the need for communication between radars.


\begin{figure}[htbp]
\centering
    \includegraphics[width=.8\textwidth]{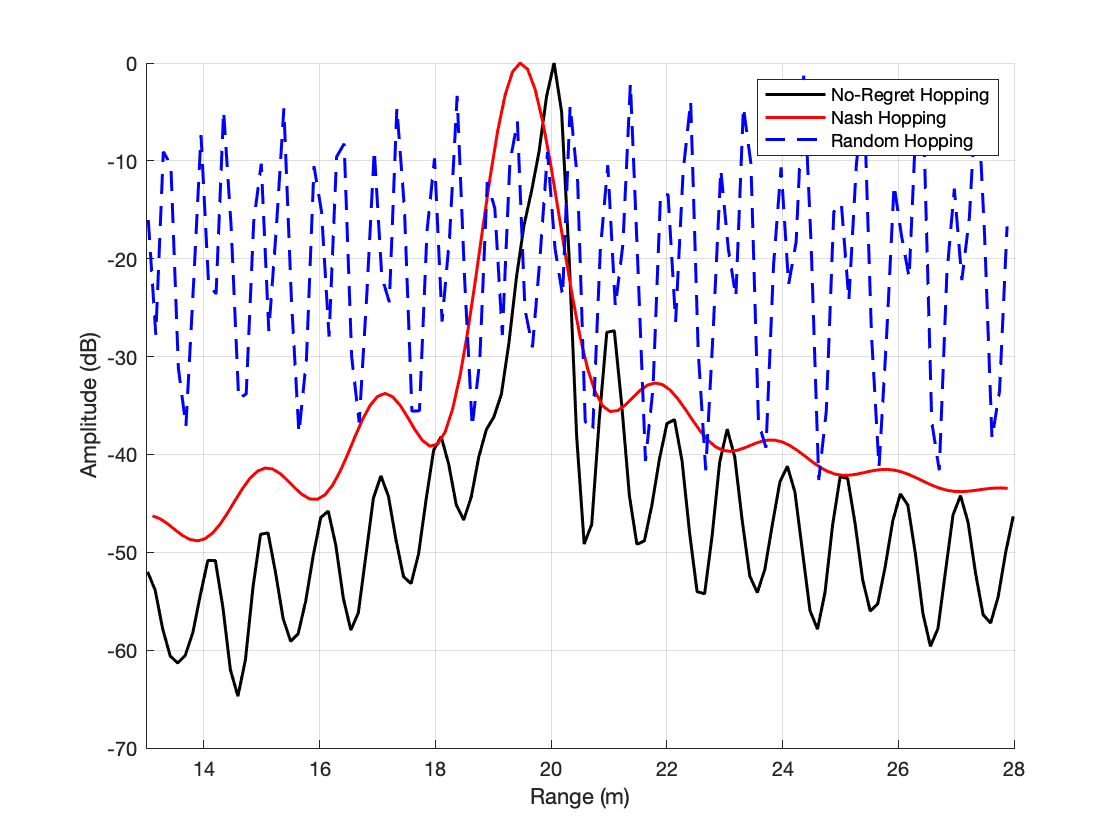}
    \caption{The range profiles at velocity $\dot{r}=-15m/s$ after applying the three strategies: \texttt{Nash Hopping}, \texttt{No-Regret Hopping}, and Uniformly Random Hopping.} \label{fig:rangereso}
\end{figure}

In Fig. \ref{fig:rangereso}, we display the range profiles at the ground truth velocity of $-15m/s$, using data generated by \texttt{Nash Hopping} and \texttt{No-Regret Hopping}. As a baseline, we include a uniformly random hopping approach, where subbands are randomly sampled with a uniform distribution over all six subbands. Although the uniformly random hopping theoretically can achieve the highest range resolution by utilizing all six subbands, it suffers from a high noise floor and low SINR due to accidental interference. Both \texttt{Nash Hopping} and \texttt{No-Regret Hopping} significantly improve SINR by effectively avoiding interference, as demonstrated in Fig. \ref{fig:main1}. Moreover, \texttt{No-Regret Hopping} achieves superior range resolution and accuracy compared to \texttt{Nash Hopping} by utilizing three subbands rather than a single subband, allowing for more efficient use of the available spectrum.



\section{Conclusion}

In this work, we pioneered the application of game-theoretic approaches to mitigate radar-to-radar interference in FMCW automotive radars.
By modeling radars as players and leveraging SINR to estimate their utility functions, we propose two frequency hopping strategies: \texttt{Nash Hopping} and \texttt{No-Regret Hopping}. The proposed \texttt{No-Regret Hopping} algorithm, designed to adaptively learn a Coarse Correlated Equilibrium (CCE), demonstrated effective interference avoidance in a decentralized manner, maintaining relatively high range resolution compared to Nash Hopping. Future work will investigate the performance of these algorithms in nonuniform radar setups and enhance their efficiency to meet real-time processing demands.

\bibliographystyle{abbrv}
\bibliography{ref}

\end{document}